\documentclass[%
 reprint,amsmath,amssymb,aps]{revtex4-1}
                                                                                                                                                                                                                                                                                                                                                                                                                                                                                                                                                                                                                                                                                                                                                                                   \usepackage{graphicx}
\usepackage{dcolumn}
\usepackage{bm}
\usepackage{braket}
\usepackage{epstopdf}
\usepackage[sort&compress]{natbib}
\begin{document}
\title{Two-atom quantum gate in hybrid cavity optomechanics}
\author{Anil Kumar Chauhan}
 \email{anilkm@iitrpr.ac.in}
\author{Asoka Biswas}%
\affiliation{%
Department of Physics, Indian Institute of Technology Ropar, Rupnagar, Punjab - 140001, India
}%
\date{\today}%

\begin{abstract}
Tracing the dynamics of a quantum system using a mesoscopic device is an important topic of interest nowadays. Here we show how a mesoscopic mechanical oscillator steers the dynamics of a coupled two-atom system and gives rise to a two-qubit SWAP gate. We have theoretically studied a generic hybrid atom-optomechanical system where two identical atoms in $\Lambda$ configuration are trapped inside the cavity and the cavity mode mediates the interaction between the atoms and the mechanical oscillator.  Adiabatic elimination of the lossy channels is adopted which in turn gives rise to an effective Hamiltonian that is responsible for a two-atom SWAP gate controlled by the mechanical motion of the oscillator. The validity of the proposal for successful implementation is assessed using presently available experimental parameters. 
\end{abstract}

\maketitle
\section{Introduction}
Quantum logic gate is one of the key elements in a quantum computer. The two-qubit gates along with a few single-qubit operations make the necessary building block in quantum computing. Such gates have been extensively studied and implemented in several physical systems, e.g., cavity quantum electrodynamics (QED) \cite{qed}, trapped ion \cite{ion}, nuclear magnetic resonance \cite{nmr}, and superconducting Cooper pairs \cite{cooper}.  An array of qubits (like in spin chain) also poses as a suitable platform for scalable quantum computing \cite{spin}, while linear optical system based on single photons has been found to be suitable for quantum communication over a long distance \cite{photon}. 

In all the existing proposals, the interaction between the qubits is often simulated by the aid of an auxiliary quantum system. For example, the quantum controlled NOT gate between two atoms, trapped inside a cavity, is obtained by their interaction with the common cavity mode. In this case, these modes are modelled as a quantum harmonic oscillator, confined to its lowest eigenstates. In this paper, we show that the interaction between two qubits can be mediated by a mesoscopic mechanical oscillator, motion of which is treated classically. Such motion-induced dynamics can give rise to a quantum logic gate. This further paves the way for controlling the quantum dynamics by mesoscopic systems. In this paper, we choose a cavity optomechanical system to demonstrate this main idea.  

A cavity optomechanical set up consists of a mechanical oscillator, coupled to the cavity mode through radiation pressure \cite{opto-review}. Such a system can be an ideal platform to  investigate the interface of classical and quantum domain, through probing the dynamics of the oscillator by the cavity mode. In fact, a mechanical oscillator of mesoscopic size and with a fundamental oscillation frequency $\omega_m$ can be an interesting object to observe both the quantum and classical effects, depending upon its ambient temperature $T$.  For example, for $T\gg T_Q=\hbar\omega_m/k_B$, the motion of the oscillator can be explained in terms of laws of classical mechanics \cite{pathria}. 
On the other hand, for $T<T_Q$, the oscillator can be prepared in a non-classical state, e.g., a squeezed state \cite{squeeze} or a cat state \cite{cat}. It can be cooled to the ground state by using pulses \cite{coolpulse} or feedback \cite{coolfeed}. The quantum effect like optomechanically induced transparency can also be obtained using an oscillator \cite{oit}. However, most of the recent research in such a system focusses on the properties of the mechanical oscillator, as an effect of external manipulation of properties of the cavity mode. In this paper, we consider a hybrid scenario, in which two atoms trapped inside the cavity can be used to perform quantum two-qubit gate, controlled by the motion of the mechanical oscillator. The cavity mode only dispersively mediates the interaction between the atoms and the oscillator. 

Specifically, we consider two atoms, each with two near-degenerate ground states and one excited state (the $\Lambda$ configuration), dispersively interacting with a cavity mode in its optomechanical set up. The excited state is adiabatically eliminated so that the ground states of each atom constitute an effective qubit. Such a system has been previously proposed towards implementing quantum logic gates in cavity QED systems \cite{asoka}. Next, the cavity mode is also adiabatically eliminated so that the effective interaction between the two qubits is governed by the 'mechanical oscillator' dynamics. The radiation pressure coupling between the cavity mode and the oscillator is generally very weak, compared to the oscillator frequency. Despite this fact, we show that the effective interaction strength between the qubits can be arbitrarily enhanced by suitably choosing the relevant detunings. Note that the atomic qubits comprise of the ground states and thereby remain unaffected by the spontaneous emission. Further, as the cavity mode is adiabatically eliminated, its decay also does not influence the gate operation. The gate can be implemented much faster than the decay of the oscillator. This makes our model immune to all sources of decoherence. We emphasize that this model is quite generic and can be extended to the different kinds of optomechanical system as well. 

The paper is structured as follows. In Sec II, we describe the model and the derive the effective Hamiltonian for the quantum logic gate. We discuss in Sec. III how the logic gate, particularly the swap gate, can be implemented using this effective Hamiltonian. In Sec. IV, we conclude the paper, with an analysis of the feasibility of this gate operation with the presently available technology.

\section{Model}
We consider a generic model of an optomechanical system, in which the cavity mode interacts with the mechanical oscillator mode through radiation pressure. One of the cavity mirrors can be chosen as the oscillator leading to a linear coupling  between the two modes \cite{gsa}. Alternatively, one could also choose a quadrature coupling, as in the case of the membrane-in-the-middle setup \cite{Thompson2008}. To keep the interaction generic, we choose the relevant interaction Hamiltonian as
\begin{equation}
H_{\rm cm}=g'a^\dag a(b+b^\dag)^n\;,
\label{genint}
\end{equation}
where $a$ and $b$ represent the annihilation operators for cavity mode and the oscillator mode, respectively. Here $g'=x_0^n\frac{\partial ^n\omega_c}{\partial x^n}$ is the relevant coupling constant, where $x_0$ is the amplitude of oscillation and $n$ represents the order of cavity-oscillator coupling. 

We next consider that two identical atoms with three relevant energy levels $|g\rangle$, $|f\rangle$, $|e\rangle$ in $\Lambda$-configuration are magneto-optically trapped inside the cavity and are interacting with the same cavity mode. The $|f\rangle\leftrightarrow |e\rangle$ transition is driven by a classical pump field with frequency $\omega_p$ and the Rabi frequency $\Omega$, while the cavity mode $a$ drives the $|f\rangle\leftrightarrow |e\rangle$ transition. The relevant Hamiltonian of the atom-cavity system can be written as
\begin{eqnarray}
H_{\rm ac}& = &\left[\Omega e^{- i \omega_p t} \sum_{i=1}^{2}\ket{e_i} \bra{g_i} + \sum_{i=1}^{2} g_{i} \ket{f_i}\bra{e_i} a^\dag +{\rm h.c.}\right]\;,\\
H_0^{\rm at} & =&  \sum_{i=1}^{2}\Big[\omega_{eg}^i \ket{e_i}\bra{e_i}+ \omega_{fg}^i \ket{f_i}\bra{f_i} \Big]\;,
\end{eqnarray}
where $g_i$ ($i\in 1,2$) are the atom-cavity coupling constants, $\omega_{\alpha g}^i$ ($\alpha \in e,f$) is the frequency difference between the levels $|\alpha_i\rangle$ and $|g_i\rangle$ of the $i$th atom, and $H_0^{\rm at}$ is the unperturbed Hamiltonian of the two atoms. 

The unperturbed Hamiltonian of the joint system  can be written as 
\begin{equation}
H_0=H_0^{\rm at}+\omega_ca^\dag a +\omega_mb^\dag b\;,
\end{equation}
where $\omega_c$ is the  cavity mode frequency and $\omega_m$ represents the frequency of the oscillator.  The cavity mode is driven by a pump field with amplitude $\epsilon$ and frequency $\omega_l$. In the reference frame, rotating with the pumping laser frequency $\omega_l$, one can obtain an effective Hamiltonian using the transformation
$H_{\mbox{rot}}=RHR^\dag + i \hbar \frac{\partial R}{\partial t} R^\dag$, where $R= \exp\{i\omega_{l} a^\dag at\}$. The total Hamiltonian then takes the following form:
\begin{equation}
H^{(1)}  =  H^{(1)}_{0}+H^{(1)}_{\rm ac}+H_{\rm cm}+H^{(1)}_{\rm pump}\;,
\end{equation}
where
\begin{eqnarray}
H^{(1)}_{0}& = & \delta a^\dag a +  \omega_{m} b^\dag b +H_0^{\rm at} \;,\nonumber\\
H^{(1)}_{\rm ac} & = &\Big[\Omega e^{-i \omega_{p} t} \sum_{i=1}^{2}\ket{e_i} \bra{g_i}\nonumber\\
&&+ \sum_{i=1}^{2} g_{i}\ket{f_i}\bra{e_i} a^\dag e^{-i \omega_{l}t}+ {\rm h.c.}\Big]\;,\nonumber\\
H^{(1)}_{\rm pump}& = &\epsilon(a+a^\dag)\;.
\end{eqnarray}
where $\delta = \omega_{c}-\omega_{l}$ is the cavity-pump detuning.

In the interaction picture with respect to the unperturbed Hamiltonian $H_0^{\rm at}$ of the atom, the Hamiltonian further reduces to 
\begin{equation}
H^{(2)} = H_0^{(2)}+H_{\rm ac}^{(2)}+H_{\rm cm}+H_{\rm pump}^{(1)}\;,
\end{equation}
where
\begin{eqnarray}
H_0^{(2)}&=&\delta a^\dag a+\omega_mb^\dag b\;,\nonumber\\
H_{\rm ac}^{(2)}&=&\Big[\Omega e^{i\Delta_{1}t}  \Big(\sum_{i=1}^2\ket{e_i} \bra{g_i} \Big)+ {\rm h.c.}\Big] \nonumber\\
& & + \Big[\sum_{i=1}^2g_{i} \ket{f_i} \bra{e_i} e^{-i( \Delta_{2}+\delta)t} a^\dag+{\rm h.c.}\Big]\;,
\end{eqnarray}
 where $\Delta_{1}=\omega_{eg}-\omega_{p}$ and  $\Delta_{2}=\omega_{ef}- \omega_{c}$ are the detunings of the classical field and cavity mode with respect to the corresponding single-photon transition for each atom. 
\begin{figure}[!h]
\includegraphics[width=0.7\linewidth]{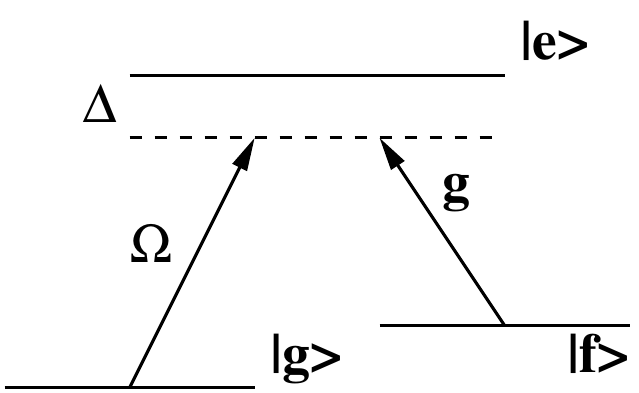} 
\caption{The level configuration for the three-level atom with the excited state $\ket{e}$ and the ground states $\ket{g}$ and $\ket{f}$. $\Omega$ and $g$ define the Rabi frequencies of the classical field and the cavity field, respectively.}
\end{figure}
%
%
 
\subsection{Effective Hamiltonian}
Next we consider that the classical field and the cavity mode have equal single-photon detuning, i.e., $\Delta_1=\Delta_2=\Delta$, pertaining to a two-photon (Raman) transition $|g\rangle\leftrightarrow |f\rangle$. In the large single-photon detuning limit, $\Delta \gg \Omega, g_{1},g_2$,  the excited states of both the atoms can be eliminated adiabatically \cite{asoka} and each three-level atom can be approximated as a two-level system (or a qubit) comprising of the ground states $\ket{g}$ and $\ket{f}$. Therefore, the Hamiltonian can be written as 
\begin{equation}
 H^{(3)} =  \delta a^\dag a + \omega_{m}b^\dag b + H_{\rm eff} +g^{\prime} a^\dag a(b+b^\dag)^n 
  + \epsilon(a+ a^\dag)\;, 
  \label{h3}
\end{equation}
where
\begin{eqnarray}
H_{\rm eff} & = &  -2\frac{|\Omega|^2}{\Delta}\ket{g_{1}g_{2}}\bra{{g_{1}g_{2}}} -2\Big[ \delta-2\frac{|g|^2}{\delta-\Delta}\Big]\ket{f_{1}f_{2}}\bra{f_{1}f_{2}} \nonumber\\
&&-\Big[ \delta -\frac{|\Omega|^2}{\delta-\Delta}+\frac{|g|^2}{\Delta} \Big] (\ket{f_{1}g_{2}}\bra{f_{1}g_{2}}+\ket{g_{1}f_{2}}\bra{g_{1}f_{2}})   \nonumber\\
&&- \frac{g \Omega }{\Delta}\Big[ \ket{g_{2}}\bra{g_{2}}\sigma_-^{(1)} a+ \ket{g_{1}}\bra{g_{1}}\sigma_-^{(2)}a+{\rm h.c.}\Big]\nonumber\\
& & + \sqrt{2} \frac{g\Omega }{\delta-\Delta}\Big[\ket{f_{2}}\bra{f_{2}}\sigma_-^{(1)} a+\ket{f_{1}}\bra{f_{1}}\sigma_-^{(2)}a+{\rm h.c.}\Big] \;,\nonumber\\
\end{eqnarray}
where $\sigma_-^{(j)}=|g_j\rangle\langle f_j|$ ($j\in 1,2$) represents a single-qubit operator and we have chosen $g_1=g_2=g$. The first three terms denote the Stark shifts of the joint states of the two atomic qubits. 

To obtain an effective coupling between the two atoms, mediated by their coupling to the cavity mode, we next consider that the cavity mode is large detuned from the cavity pump field such that the cavity is not sufficiently populated with photons. We first obtain the Heisenberg equation of motion of the cavity mode $a$ as
\begin{eqnarray}
 \dot{a} & = & -i[a,H^{(3)}]\nonumber\\
 &=&-i\Big[\delta a -\frac{\Omega^* g^*}{\Delta}\Big(\ket{g_{2}}\bra{g_{2}}\sigma_+^{(1)}+\ket{g_{1}}\bra{g_{1}}\sigma_+^{(2)}\Big)\nonumber\\
 & &+\sqrt{2}\frac{\Omega^* g^*}{\delta-\Delta}\Big(\ket{f_{2}}\bra{f_{2}}\sigma_+^{(1)}+\ket{f_{1}}\bra{f_{1}}\sigma_+^{(2)}\Big) \nonumber \\
 &&+ g^{\prime}a^\dag a(b+b^{\dag})^n +\epsilon \Big]\;.
\end{eqnarray} 
In the limits $\delta \gg \frac{\Omega g}{\Delta}, g^{\prime}$, we can adiabatically eliminate the cavity mode $a$ by substituting $\dot{a}\approx 0 $. This results in the following operator identity in the limit $\delta \gg \epsilon$:
\begin{eqnarray}
a &\approx &  \frac{1}{\delta}\Big[\frac{\Omega^* g^*}{\Delta}\Big(\ket{g_{2}}\bra{g_{2}}\sigma_+^{(1)}+\ket{g_{1}}\bra{g_{1}}\sigma_+^{(2)}\Big)\nonumber\\
& & -\sqrt{2}\frac{\Omega^* g^*}{\delta-\Delta}\Big(\ket{f_{2}}\bra{f_{2}}\sigma_+^{(1)}+\ket{f_{1}}\bra{f_{1}}\sigma_+^{(2)}\Big)\Big]\;. 
\end{eqnarray}
It is important to note that we have neglected the cavity pump field $\epsilon$ in the above equation. This means that the cavity photon numbers do not change during the time-evolution, i.e., $d(a^\dag a)/dt$ remains negligibly small \cite{anil2016}. This leads to an effective coupling between the atoms through exchange of virtual photons only. 

Upon using the above identity in Eq. (\ref{h3}),  the Hamiltonian then takes the form
\begin{eqnarray}
H^{(4)} & = & \Big[ -A\ket{g_{1}g_{2}}\bra{g_{1}g_{2}}-C\ket{f_{1}f_{2}}\bra{f_{1}f_{2}}  \nonumber\\
&&-B (\ket{g_{1}f_{2}}\bra{g_{1}f_{2}}+\ket{f_{1}g_{2}}\bra{f_{1}g_{2}})\nonumber\\
& & - D\Big(\sigma_+^{(1)}\sigma_-^{(2)}+\sigma_-^{(1)}\sigma_+^{(2)}\Big) +\omega_{m}b^\dag b \Big]\;,
\label{h4}
\end{eqnarray}
where
\begin{eqnarray}
A& =& 2\frac{|\Omega|^2}{\Delta}-2\left |\frac{g\Omega}{\delta\Delta}\right |^2g^{\prime}(b+b^\dag)^n\;,\nonumber\\  
B & = &\frac{1}{\delta}\left |\frac{\Omega g}{\Delta}\right |^2-2\left |\frac{\Omega g}{\delta(\delta-\Delta)}\right |^2g^{\prime}(b+b^\dag)^n + \Big[\delta-\frac{|\Omega|^2}{\delta-\Delta}+ \frac{|g^2|}{\Delta} \Big]\;,\nonumber \\
C & = & \frac{4}{\delta}\left |\frac{\Omega g}{\delta-\Delta}\right |^2+ 2\Big[\delta-2\frac{|g|^2}{\delta-\Delta} \Big] \;,\nonumber\\ 
D&=&\frac{1}{\delta}\left |\frac{\Omega g}{\Delta}\right |^2-2\left |\frac{\Omega g}{\delta(\delta-\Delta)}\right |^2g^{\prime}(b+b^\dag)^n \;.
\label{coeff}
\end{eqnarray}
 
To explore the possibility of the atomic swap gate, we choose the Hilbert subspace of the atomic states $\ket{g_{1}f_{2}}$ and $\ket{f_{1}g_{2}}$ only.
In the large detuning regime $\Delta \gg \Omega,g$ we assume that $\delta \approx \Delta$ such that $\delta - \Delta =\xi \rightarrow 0$. In this limit,  we retain only the terms, relevant to the above subspace and retain the leading terms  in the coefficients $B$ and $D$ in (\ref{coeff}). The above Hamiltonian therefore can be further reduced to the following form:
\begin{equation}{\label{inter}}
H^{(5)} =  \left(-B+\omega_mb^\dag b\right)\mathbf{\hat{1}}_{\rm atom}
- D\Big(\sigma_+^{(1)}\sigma_-^{(2)}+ \sigma_-^{(1)}\sigma_+^{(2)}\Big) \;,
\end{equation} 
where 
\begin{equation}
B  \approx  \delta-\frac{|\Omega|^2}{\xi}+ \frac{|g^2|}{\Delta} \;\;,\;\;   D\approx  -2\left |\frac{\Omega g}{\delta\xi}\right |^2g^{\prime}(b+b^\dag)^n\;,
 \end{equation}
 and $\mathbf{\hat{1}}_{\rm atom}=\ket{g_{1}f_{2}}\bra{g_{1}f_{2}}+\ket{f_{1}g_{2}}\bra{f_{1}g_{2}}$ is the identity operator in the atomic subspace.
In the interaction picture with respect to the oscillator Hamiltonian [the first term in (\ref{inter})], the Hamiltonian of the atom-oscillator system can then be written as
\begin{equation} \label{nu1}
V_{\rm eff}  =  \eta X'\Big(\sigma_+^{(1)}\sigma_-^{(2)}+\sigma_-^{(1)}\sigma_+^{(2)}\Big) \;,
\end{equation}
where 
\begin{equation}
\eta=2(\sqrt{2})^n\left|\frac{\Omega g}{\delta\xi}\right|^2g^\prime\;\;\;,\;\;\;X'=e^{i\omega_mtb^\dag b}X^ne^{-i\omega_mtb^\dag b}\;.
\end{equation}
The operator $X'$ contains a position quadrature term $X=(b+b^\dag)/\sqrt{2}$, rotated by a time-dependent phase $\omega_mtb^\dag b$, relevant to the motion of the mechanical oscillator. At large temperature $T\gg T_Q$, the thermal effect dominates over the quantum nature of the oscillator and its motion can be treated as that of a classical harmonic oscillator, the dynamics of which can then be described as  $X_{\rm cl}(t)=X_{\rm cl}(0)\cos(\omega_{m}t)$. Here $X_{\rm cl}(0)$ is the dimensionless amplitude of the classical oscillation, normalized by the zero-point uncertainty of position $x_{\rm zpf}$ and we have assumed that the oscillator starts its motion from the rest. Therefore, the effective two-atom Hamiltonian finally reduces to 
\begin{equation}
V_{\rm eff}(t)  = \lambda \cos^n(\omega_{m}t)\Big[\sigma_+^{(1)}\sigma_-^{(2)}+\sigma_-^{(1)}\sigma_+^{(2)}\Big]\;,
\label{vfinal}
\end{equation}
where
\begin{equation}
 \lambda  = 2(\sqrt{2})^n\left |\frac{\Omega g}{\delta\xi}\right |^2 g^{\prime} X(0)^n\;.
\end{equation}
The above form of the effective coupling strength of $\lambda$ suggests that it can be arbitrarily enhanced by choosing $\xi\rightarrow 0$, i.e., by suitably choosing the detunings $\delta$ and $\Delta$ of the cavity mode with the pump field and the atomic transition, respectively. In the following, we choose $X(0)=1$.
\section{Quantum swap gate}
Clearly, the above Hamiltonian (\ref{vfinal}) is of the form $\vec{\sigma}_1.\vec{\sigma}_2-\sigma_{1z}\sigma_{2z}$, which is known to be the operator for a two-qubit swap gate \cite{loss}.  
It  indicates that the two-atom logic gates can be implemented with the help of the motion of the oscillator. The sinusoidal driving field, generated by this motion, leads to a Rabi-like oscillation between the relevant energy levels of the two atoms. To see this, we choose the two-atom basis $\ket{g_1g_2}$, $\ket{g_1f_2}$, $\ket{f_1g_2}$, and $\ket{f_1f_2}$. The states $\ket{g_{1}f_{2}}$ and  $\ket{f_{1}g_{2}}$ are coupled through the Hamiltonian (\ref{vfinal}), while the other two states remain uncoupled. Note that (\ref{vfinal}) involves only the ground states of the atoms and therefore the dynamics of the relevant states is not influenced by the spontaneous emission. The respective probability amplitudes $b_1$ and $b_2$ of these states evolve with time through the following Schr\"odinger's equations:
\begin{eqnarray}
\dot{b}_{1}=- i  \lambda \cos^n(\omega_{m}t) b_{2}\;,\nonumber\\
\dot{b}_{2}=-i \lambda \cos^n(\omega_{m}t) b_{1}\;.
\label{beq}
\end{eqnarray}
To solve, we consider two different configurations of the optomechanical system: (a) One of the cavity mirrors acts as a mechanical oscillator, leading to a linear coupling of the form (\ref{genint}) with $n=1$, (b) An oscillator is suspended at a node or an antinode of the cavity frequency inside the cavity, in which case both the cavity mirrors are kept fixed. This leads to a quadratic coupling between the cavity mode and the mirror with $n=2$ in Eq. (\ref{genint}). 

In the former case, the solutions can be analytically obtained as 
\begin{eqnarray}
b_1(t') &=&b_1(0)e^{i\lambda'\sin(t')}+b_2(0)e^{-i\lambda'\sin(t')}\;,\nonumber\\
b_2(t') &=& -b_1(0)e^{i\lambda'\sin(t')}+b_2(0)e^{-i\lambda'\sin(t')}\;,
\end{eqnarray}
where $\lambda'=\lambda/\omega_m$ and $t'=\omega_mt$ are the coupling constant and the time, respectively, normalized with respect to $\omega_m$. 
Clearly, this implements a swap gate between the two atoms as $\ket{g_1,f_2}\leftrightarrow \ket{f_1,g_2}$ at a time $T_1$ (in unit of $1/\omega_m$), given by $\sin\{\lambda'\sin[T_1]\}=\pm 1$, or
\begin{equation}
T_1=\sin^{-1}[(2s+1)\pi/2\lambda']\;,
\end{equation}
where  $s\le (\lambda'/\pi)-1/2$ is an integer. As $\lambda'$ increases, the time-scale $T_1$ of the swap gate decreases [see Fig. 2(a)]. Further, for a constant $\lambda'$, $T_1$ can have only a limited set of possible values, constrained to the upper limit of $s$.

\begin{figure}[!h]
$\begin{array}{c}
\includegraphics[width=1\linewidth]{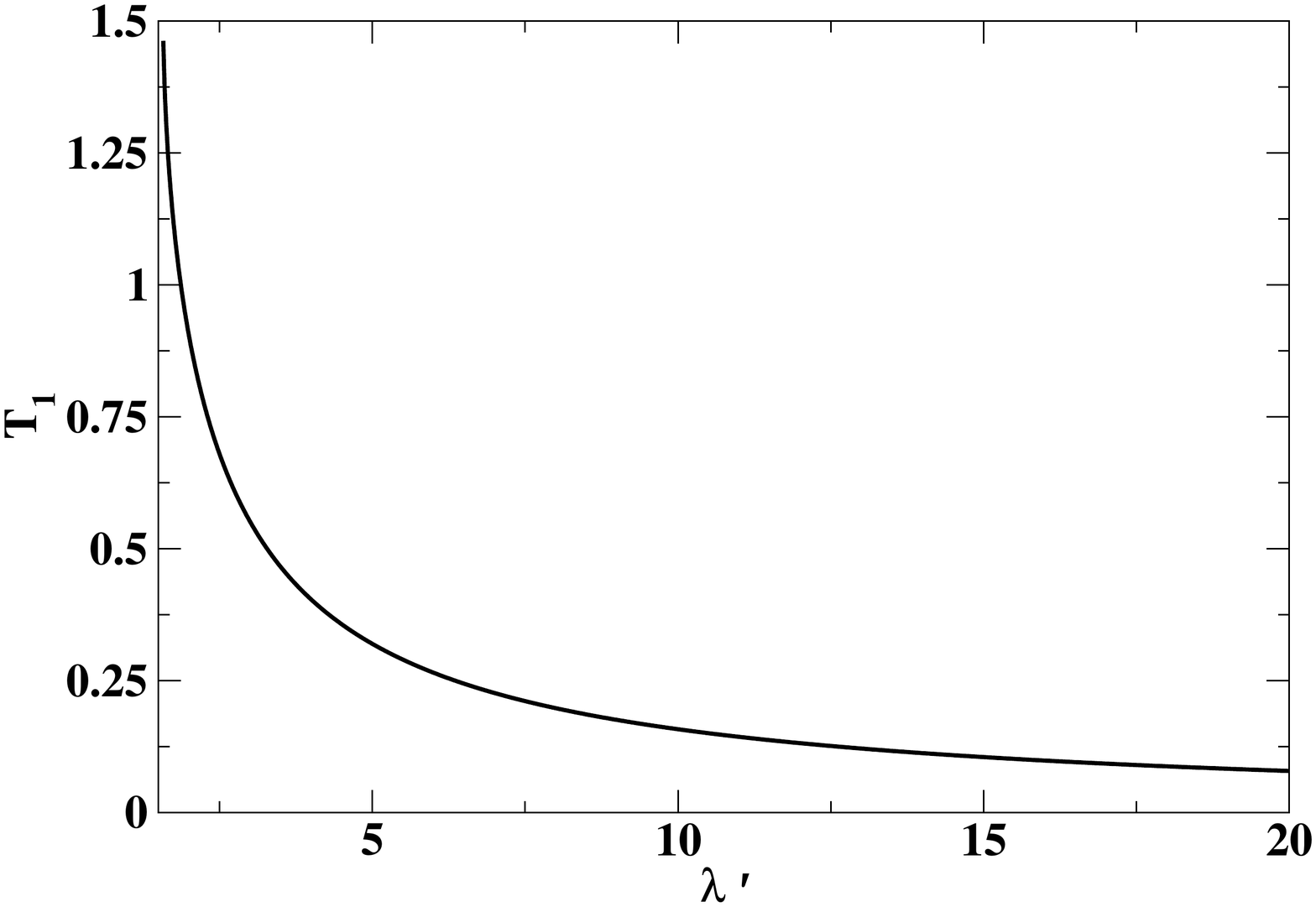} \\ \includegraphics[width=1\linewidth]{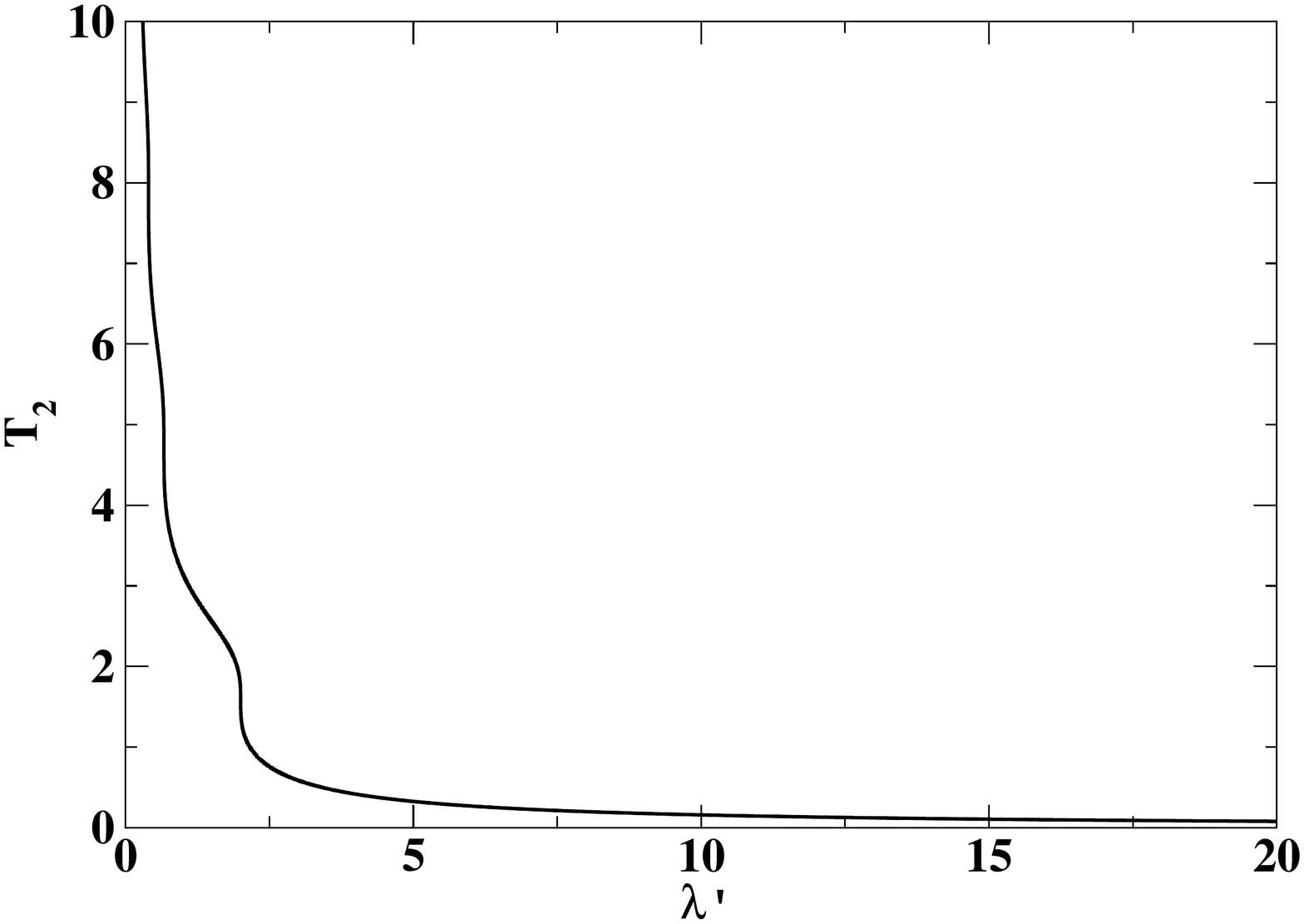}
\end{array}$
\caption{Variation of time-scale  (a) $T_1$  and  (b) $T_2$ (in unit of $1/\omega_m$) for swapping operation with the coupling constant $\lambda'=\lambda/\omega_m$. We have chosen $s=0$ corresponding to minimum time for swapping for a given $\lambda$. Note that the minimum value of $\lambda'$ in (a) is $\pi/2$.}
\end{figure}
On the other hand, for $n=2$, the Eq. (\ref{beq}) can be solved as
\begin{eqnarray}
b_{1}(t')&=&b_1(0)\exp\left[i\lambda'\left(\frac{t'}{2}+\frac{1}{4}\sin(2t')\right)\right] \nonumber\\
&&+b_2(0)\exp\left[-i\lambda'\left(\frac{t'}{2}+\frac{1}{4}\sin(2t')\right)\right]\\
b_{2}(t')&=& -b_1(0)\exp\left[i\lambda'\left(\frac{t'}{2}+\frac{1}{4}\sin(2t')\right)\right]\nonumber\\ &&+b_2(0)\exp\left[-i\lambda'\left(\frac{t'}{2}+\frac{1}{4}\sin(2t')\right)\right]
\end{eqnarray}
This will lead to a swap gate at a time $T_2$ (in unit of $1/\omega_m$), that follows the equation
\begin{eqnarray}
2T_2+\sin(2T_2)=\frac{2\pi(2s+1)}{\lambda'}\;\;\;,\;\;\;s \in 0,1,2,3,\cdots\;.
\label{t2}
\end{eqnarray}
We show in Fig. 2(b), how $T_2$, as a solution of Eq. (\ref{t2}) decreases with increase in $\lambda'$. As $\lambda'$ can be arbitrarily enhanced by choosing $\xi$ arbitrarily small, $T_2$ can also be made small. This makes the gate operation quite fast: one can obtain a time-scale of $7.87\times 10^{-2}/\omega_m$ for $\lambda=20\omega_m$, while the time-scale of decay of the oscillator can be of the order of $10^2/\omega_m$. This clearly indicates that oscillator damping does not affect the qubit dynamics.  It is interesting to note that for large $\lambda'$, the time-scale does not change substantially with increase in $\lambda'$. This means that gate operation becomes robust upon variation of the system parameters $\Omega$, $g$, $\delta$, and $\Delta$. 

\section{Conclusions}
Our model of quantum  gate can be well implemented using the available experimental parameters. For example, in a linear coupling set up ($n=1$) using a torroidal microcavity \cite{torroid}, the fundamental frequency $\omega_m$ of the oscillator can be of the order of $2\pi \times 78$ MHz (corresponding to $T_Q\sim 4$ mK). The corresponding coupling constant $g'=x_0\frac{\partial \omega_c}{\partial x}|_{x=0}$ therefore becomes $3.4\times 10^4$ Hz, where $x_0$ is the amplitude of the oscillation. We have chosen $x_0=10x_{\rm zpf}$, where $x_{\rm zpf}=\sqrt{\hbar/2m\omega_m}$ is the zero-point uncertainty of the position and $m$ is the mass of the oscillator.  Considering the Rabi frequencies for driving the atomic transitions as $\Omega=g=1$ MHz and the cavity-pump detuning $\delta\sim 10$ MHz, we can have the effective coupling for the swap gate operation as $\lambda=(2\sqrt{2}/\xi^2)\times 10^{10}$. Choosing $\xi=\delta-\Delta=2$ Hz, we have $\lambda'=\lambda/\omega_m\approx 14.42$. As seen in Fig. 2(a), this corresponds to a time for the swap gate operation $\sim 0.125/\omega_m =  2.5\times 10^{-10}$ s, which is much smaller than all the relevant decay times of the oscillator, cavity, and atomic excited states. Note that for $\xi\rightarrow 0$ and $\delta=10$ MHz, the condition $\Delta\gg \Omega, g$ for the adiabatic elimination of excited state of the atom is well satisfied. 

Further, we consider a membrane-in-the-middle setup ($n=2$) \cite{Thompson2008}, with a membrane of mass $m=40$ ng and $\omega_m=2\pi\times 134$ kHz, corresponding to the zero-point uncertainty in the position  $x_{\rm zpf}=1.24\times 10^{-15}$ m and $T_Q=6.5 \mu$K. If the membrane is placed at a node or an antinode of frequency $\omega_c$ inside the cavity, the coupling constant can be written as $g'=x_0^2\left.\frac{\partial^2\omega_c}{\partial x^2}\right |_{x=0}$. Choosing $x_0=10x_{\rm zpf}$ as before, we can have, for a typical set-up, $g'=5.65\times 10^{-5}$ Hz \cite{law}. For  $\Omega=g=1$ MHz and $\delta\sim 10$ MHz, the effective coupling for the swap gate operation becomes $\lambda=2.26\times 10^6 /\xi^2$ Hz. Therefore, this coupling constant can be enhanced to a very large value for $\xi\rightarrow 0$. For example, choosing $\xi=1$ Hz, we have $\lambda'=2.684$, corresponding to a swap gate time $\sim 0.7/\omega_m =8.3\times 10^{-7}$ s.  

In conclusion we have proposed a generic model of hybrid atom-optomechanical system to achieve a two-atom swap gate controlled by the motion of a mesoscopic mechanical oscillator. Two identical atoms in their $\Lambda$ configuration are considered to be trapped inside a cavity and driven, in their two dipole transitions, by an external laser field and the cavity mode, respectively. The cavity mode is also coupled to a mechanical oscillator through radiation pressure force. By adiabatically eliminating the cavity mode and the atomic excited states, we then derive an effective Hamiltonian between the two atoms, that essentially leads to the two-qubit SWAP gate operation. The time scale of the gate operation depends on the effective coupling strength of the oscillator with the atomic system, which can be arbitrarily enhanced by choosing the detunings of the cavity modes with the atomic transition and the  cavity pump field.  We have shown that the swap gate can be implemented using presently available technology.


\begin{thebibliography}{99}
\bibitem{qed}J. M. Raimond, M. Brune, and S. Haroche, Rev. Mod. Phys. {\bf 73}, 565 (2001). 
\bibitem{ion}D. Leibfried, R. Blatt, C. Monroe, and D. Wineland, Rev. Mod. Phys. {\bf 75}, 281 (2003).
\bibitem{nmr}L. M. K. Vandersypen and I. L. Chuang, Rev. Mod. Phys. {\bf 76}, 1037 (2005).
\bibitem{cooper}Yuriy Makhlin, Gerd Schön, and Alexander Shnirman, Rev. Mod. Phys. {\bf 73}, 357 (2001).
\bibitem{spin}Florian Meier, Jeremy Levy, and Daniel Loss, Phys. Rev. B {\bf 68}, 134417 (2003).
\bibitem{photon}Pieter Kok, W. J. Munro, Kae Nemoto, T. C. Ralph, Jonathan P. Dowling, and G. J. Milburn, Rev. Mod. Phys. {\bf 79}, 135 (2007).
\bibitem{opto-review}M. Aspelmeyer, T. J. Kippenberg, and F. Marquardt, Rev. Mod. Phys. {\bf 86}, 1391 (2014).
\bibitem{pathria}R. K. Pathria, {\it Statistical Mechanics}, 2nd Ed. (Butterworth Heinemann, Oxford, 1996).

\bibitem{squeeze}K. J\"ahne, C. Genes, K. Hammerer, M. Wallquist, E. S.
Polzik, and P. Zoller, Phys. Rev. A {\bf 79}, 063819 (2009); A. Szorkovszky, A. C. Doherty, G. I. Harris, and W. P. Bowen, Phys. Rev. Lett. {\bf 107}, 213603 (2011); A. Szorkovszky, G. A. Brawley, A. C. Doherty, and W. P. Bowen, Phys. Rev. Lett. {\bf 110}, 184301 (2013); M. R. Vanner et al., Proc. Natl. Acad. Sci., USA {\bf 108}, 16182 (2011).
\bibitem{cat}H. Tan, Phys. Rev. A {\bf 89}, 053829 (2014).
\bibitem{coolpulse}J.-Q. Liao and C. K. Law, Phys. Rev. A {\bf 84}, 053838 (2011).
\bibitem{coolfeed}S. Mancini, D. Vitali, and P. Tombesi, Phys. Rev. Lett.
{\bf 80}, 688 (1998); P.-F. Cohadon, A. Heidmann, and M. Pinard, Phys.
Rev. Lett. {\bf 83}, 3174 (1999); A. Hopkins, K. Jacobs, S. Habib, and K. Schwab, Phys. Rev. B {\bf 68}, 235328 (2003); D. Kleckner and D. Bouwmeester, Nature (London)
{\bf 444}, 75 (2006).
\bibitem{oit}G. S. Agarwal and S. Huang, Phys. Rev. A {\bf 81}, 041803 (2010); S. Weis {\it et al.\/}, Science {\bf 330}, 1520 (2010).
\bibitem{asoka}A. Biswas and G. S. Agarwal, Phys. Rev. A {\bf 69}, 062306 (2004).

\bibitem{gsa}G. S. Agarwal, {\it Quantum Optics}, (Springer, Berlin, 2015).
\bibitem{Thompson2008}J. D. Thompson, B. M. Zwickl, A. M. Jayich, F. Marquardt, S. M. Girvin, and J. G. E. Harris, Nature (London) {\bf 452}, 72 (2008).
\bibitem{anil2016}Anil Kumar Chauhan and Asoka Biswas, arXiv:1512.03900 (to appear in Phys. Rev. A).
\bibitem{loss}Daniel Loss and David P. DiVincenzo, Phys. Rev. A {\bf 57}, 120 (1998).
\bibitem{torroid}E. Verhagen, S. Deleglise, S. Weis, A. Schliesser, and T. J. Kippenberg, Nature (London) {\bf 482}, 63 (2012).
\bibitem{law}H. K. Cheung and C. K. Law, Phys. Rev. A {\bf 84}, 023812 (2011).
\end{thebibliography}
\end{document}